\documentclass[conference]{IEEEtran}
\IEEEoverridecommandlockouts
\usepackage{multicol}
\usepackage{comment}
\usepackage{cite}
\usepackage{hyperref}
\usepackage{lipsum}
\usepackage{adjustbox}
\usepackage{amsmath,amssymb,amsfonts}
\usepackage{float}
\usepackage{algorithmic}
\usepackage{pgfplots}
\usepackage{graphicx}
\usepackage{textcomp}
\usepackage{xcolor}
\usepackage{hyperref}
\usepackage{comment}

\def\BibTeX{{\rm B\kern-.05em{\sc i\kern-.025em b}\kern-.08em
    T\kern-.1667em\lower.7ex\hbox{E}\kern-.125emX}}

\newcommand{\squishlist}{
   \begin{list}{$\bullet$}
    { \setlength{\itemsep}{0pt}      \setlength{\parsep}{0pt}
      \setlength{\topsep}{-3pt}       \setlength{\partopsep}{0pt}
      \setlength{\listparindent}{-2pt}
      \setlength{\itemindent}{-5pt}
      \setlength{\leftmargin}{1em} \setlength{\labelwidth}{0em}
      \setlength{\labelsep}{0.5em} } }

\newcommand{\squishend}{
    \end{list}  }
    
\begin{document}

\title{SMART Paths for Latency Reduction in ReRAM Processing-In-Memory Architecture for CNN Inference}

\author{\IEEEauthorblockN{Sho Ko}
\IEEEauthorblockA{\textit{School of ECE} \\
\textit{Georgia Tech}\\
Atlanta, GA, USA \\
sko.45@gatech.edu}
\and
\IEEEauthorblockN{Shimeng Yu}
\IEEEauthorblockA{\textit{School of ECE} \\
\textit{Georgia Tech}\\
Atlanta, GA, USA \\
shimeng.yu@ece.gatech.edu}\\
}
\maketitle

\begin{abstract}
This research work proposes a design of an analog ReRAM-based PIM (processing-in-memory) architecture for fast and efficient CNN (convolutional neural network) inference. For the overall architecture, we use the basic hardware hierarchy such as node, tile, core, and subarray. On the top of that, we design intra-layer pipelining, inter-layer pipelining, and batch pipelining to exploit parallelism in the architecture and increase overall throughput for the inference of an input image stream. We also optimize the performance of the NoC (network-on-chip) routers by decreasing hop counts using SMART (single-cycle multi-hop asynchronous repeated traversal) flow control. Finally, we experiment with weight replications for different CNN layers in VGG (A-E) for large-scale data set ImageNet. In our simulation, we achieve 40.4027 TOPS (tera-operations per second) for the best-case performance, which corresponds to over 1029 FPS (frames per second). We also achieve 3.5914 TOPS/W (tera-operaions per second per watt) for the best-case energy efficiency. In addition, the architecture with aggressive pipelining and weight replications can achieve $14\times$ speedup compared to the baseline architecture with basic pipelining, and SMART flow control achieves $1.08\times$ speedup in this architecture compared to the baseline. Last but not least, we also evaluate the performance of SMART flow control using synthetic traffic.
\end{abstract}


\section{Introduction}
Recently, artificial intelligence (AI) and machine learning (ML) techniques have become more and more mature. AI/ML gradually become an indispensable part of our life. Neural network (NN) is an important category of AI/ML that mimics how human thinks. Neuromorphic computing has gained more and more popularity in both academia and industry. Specifically, CNN is one type of NN that has revolutionized deep learning applications by achieving unprecedented accuracy in computer vision and pattern recognition applications.

From the software perspective, the data flow in a CNN is clear and simple. Programming a CNN in a certain deep learning framework can be easily done. However, from the hardware perspective, CNN is extremely computation intensive and power hungry. The reason is that CNN requires billions of MAC (multiply and accumulate) operations, which quickly ties up a conventional CPU or even a GPU. Therefore, the circuit and architecture research communities gradually focus on designing efficient digital accelerator for CNNs. The building blocks of these accelerators are still transistors and logic gates. Researchers are trying to use the basic modules to design digital circuit and computer architecture that compute CNNs most efficiently. However, as the digital processor reaches the upper limit of its computing capability, researchers are looking for the next-generation hardware solution. Recently, there have been some emerging eNVMs (embedded non-volatile memories) designed from the device perspective. As these emerging technologies become more and more mature, people are researching on using them to build efficient PIM circuit and architecture for processing CNNs. Our paper presents an efficient ReRAM-based PIM architecture for CNN inference. The contributions are summarized in the following three bullet points.

\begin{enumerate}
   \item We utilize one type of eNVM, ReRAM (resistive random access memory) to design a PIM architecture for fast and efficient CNN inference.
   \item We optimize the architecture from both the processing side and the interconnect side. From the processing side, we design intra-layer pipelining, inter-layer pipelining, and batch pipelining to exploit parallelism in the architecture and increase overall throughput for the inference of an input image stream. From the interconnect side, we maximizes the performance of the NoC routers by decreasing hop counts using SMART flow control. In addition, we experiment with weight replications to further accelerate the architecture.
   \item  To report throughput and energy efficiency,  we run the cycle-accurate simulation for the processing side by building a C++ simulator from scratch, and we use the garnet2.0 simulator for the interconnect side. Our design achieves great speedup compared to the baselines. Independently, we also evaluate the performance of SMART flow control using synthetic traffic in garnet2.0.
\end{enumerate}

\section{Background}
\subsection{CNN Training and Inference}

\begin{figure}[H]
\centering
\vspace{-10pt}
\includegraphics[width=85mm]{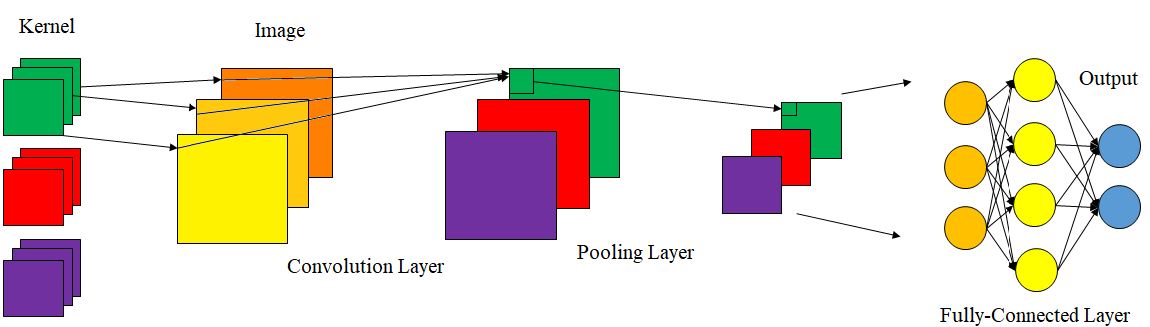}
\caption{CNN Inference.}
\vspace{-10pt}
\label{cnn}
\end{figure}

In this section, we present the CNN algorithm implemented in our design. A CNN has two phases: training and inference. To start with, the weights in a CNN are initialized with random values. Then the training process will update and refine the weights to a specific data set. Finally, a well-trained a CNN can be used for inference of new images. Typically, the training process has much more power and time consumption than the inference process because training requires forward propagation, back propagation, and weight update while inference only requires forward propagation. However, a CNN needs to be trained only once, and then it can be used for inference for many times. For the inference process of a CNN, it consists of multiple layers with three basic types: convolution layers, pooling layers, and fully-connected layers, as shown in Fig. \ref{cnn}. In this paper, we focus on the inference process of a CNN.

\subsection{ReRAM Device}
From the device perspective, resistance-based eNVMs have become more and more mature and manufacturable. Technologies such as ReRAM \cite{ReRAM}, PCM (phase change memory) \cite{PCM}, and STT-MRAM (spin-transfer torque magnetic random access memory) \cite{STT-MRAM} start to gain more and more popularity. These eNVMs have much smaller cell size than SRAM. They can also achieve MLC (multiple bits per cell). Therefore, they can be utilized to map the entire weights on-chip at once and eliminate off-chip accesses. They are also non-volatile and CMOS-process compatible. Their access speed is within $10$ ns, which is in the same magnitude as SRAM.

\subsection{Analog PIM Circuit}

\begin{figure}[t!]
\centering
\includegraphics[width=60mm]{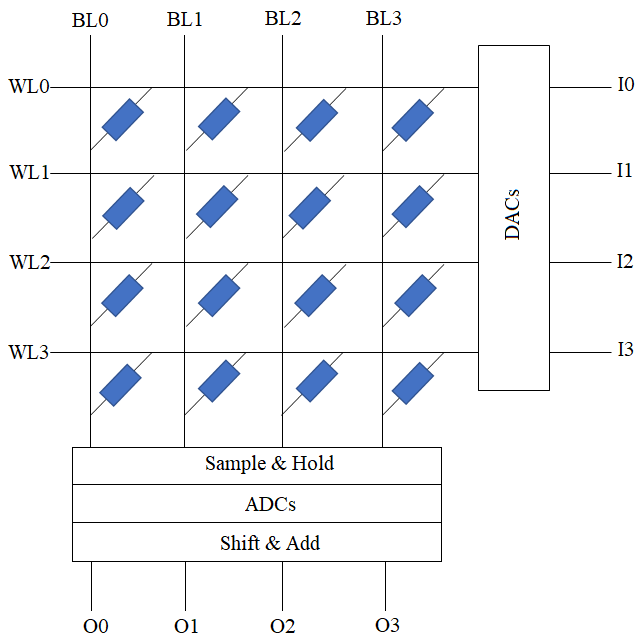}
\caption{ReRAM Array with Peripheral Circuits.}
\vspace{-15pt}
\label{rram}
\end{figure}

From the circuit perspective, 2D ReRAM is a grid structure consisting of multiple ReRAM cells, as shown in Fig. \ref{rram}. Such design can exploit the analog characteristics of ReRAM to perform fast and energy-efficient matrix multiplication and convolution. Vector-matrix multiplication can be easily calculated using ReRAM, because of two basic electrical theorems, Ohm's law and Kirchhoff's current law. Ohm's law states that the current through a resistor is equal to the voltage across the resistor divided by the resistance of the resistor ($I=V/R$), which is also equal to the voltage across the resistor multiplied by the conductance of the resistor ($I=VG$). This law makes performing analog floating-point multiplication possible. Kirchhoff's current law states that the total current output is equal to the sum of all input current for a node in the circuit. This law makes performing analog floating-point addition possible. Vector-matrix multiplication can be mapped to ReRAM in the following three steps. First, the digital input is converted to analog signals by DACs (digital-to-analog converters) and then mapped to the voltage on horizontal WLs (word lines); Second, the weight matrix is quantized and then mapped to the conductance of ReRAM cells; Third, the analog output signals are read from the current on the vertical BLs (bit lines), stored in sample \& hold units, converted to digital output by ADCs (analog-to-digital converters), and some columns are shifted and added together to produce the final results.

\subsection{PIM Architecture}
Recently, several ReRAM-based PIM architectures have been presented for CNN inference, such as PUMA \cite{PUMA}, ISAAC \cite{ISAAC}, and PRIME \cite{PRIME}. 

PUMA creates its own compiler and domain-specific ISAs (instruction set architectures) to make the architecture more general-purpose, programmable, and reconfigurable. It's a spatial architecture in which each tile is executing its own ISAs simultaneously with all other tiles. It uses a state machine to synchronize among different cores, it has a large synchronization overhead. In addition, the penalty of ISA, instruction decoder, and instruction memory is also large if the workloads are only CNNs.

Unlike PUMA, ISAAC and PRIME are ASICs specifically for CNN inference. PRIME is slightly different from ISAAC in the sense that PRIME stores positive and negative weights in separate subarrays while ISAAC stores them in the same subarray and uses a small trick to differentiate based on the MSB of a positive 2's complementary number is 0 while the MSB of a negative 2's complementary number is 1. Therefore, PRIME comes with more area and power penalty, which leads to less area and energy efficiency.

\section{Overall Architecture}
\begin{figure}[t!]
\centering
\includegraphics[width=85mm]{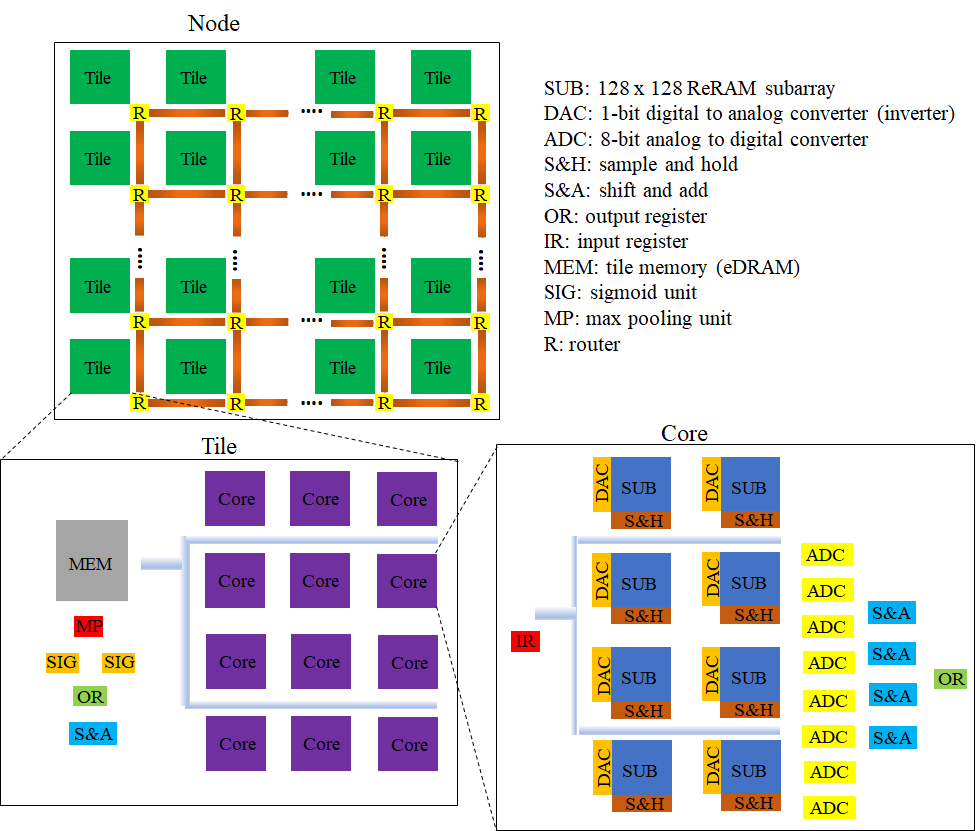}
\caption{Overall Architecture of a Node.}
\vspace{-15pt}
\label{node}
\end{figure}

The overall chip, also called a node, as shown in Fig. \ref{node}. The node is composed of $16 \times 20 = 320$ tiles. Each tile has a outer associated with it. The routers form a mesh structure. Within each tile, there are 12 cores, a local memory of 64KB eDRAM, a shift \& add unit, an output register of 2KB eDRAM, two sigmoid units, and a max pooling unit. Within each core, there are eight ReRAM subarrays of size $128 \times 128$, $128 \times 8$ 1-bit DACs, $128 \times 8$ sample \& hold units, eight ADCs with 8-bit resolution, four shift \& add units, an input register of 2KB eDRAM, and an output register of 2KB eDRAM. There are buses within each tile and each core. The number of each component is designed so that there is no structural hazard during run time. For our design, the weights and feature maps are both fixed 16 bits. Lots of previous research has shown that 16 bits are accurate enough for CNN inference. We conservatively assume 2-bit MLC for each ReRAM cell. Therefore, we need eight cells across eight different columns to encode all of them. In addition, our DAC is of 1-bit resolution, which is trivial. Since 16-bit DAC has too much noise and takes too much area and power, we choose to pass in the 16-bit IFM bit by bit sequentially in 16 cycles. Therefore, we only need 1-bit DACs. Note that since we partition the weight spatially across different columns and we also partition the input temporally within the same column, the shift and add unit after the ADC will be necessary to produce the correct final results.

\begin{figure}[t!]
\centering
\includegraphics[width=85mm]{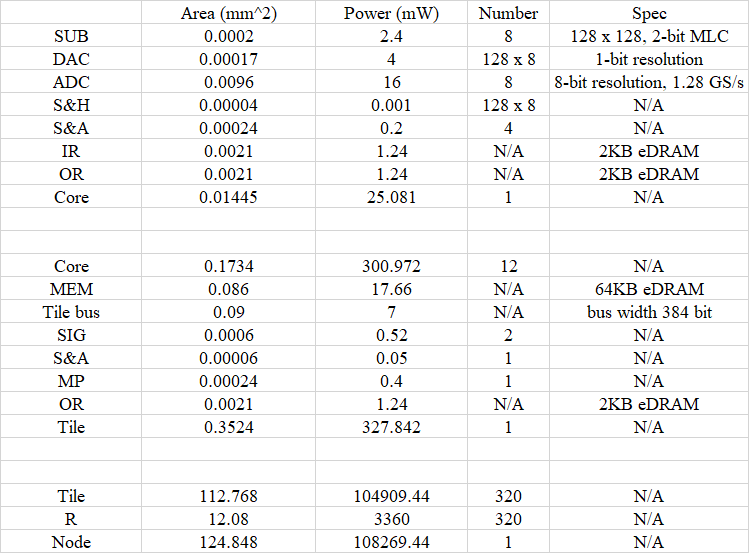}
\caption{Power and Area of Each Hardware Component.}
\vspace{-15pt}
\label{power}
\end{figure}

Fig. \ref{power} shows the power and area of each individual component, we gather the data from PUMA \cite{PUMA} and ISAAC \cite{ISAAC}, both of which are in 32 nm CMOS technology node. Note that this table shows the power consumption when the component is functioning. The node has a total area of 124.848 $mm^2$. The total power is 108.26944 W, which is the peak power consumption assuming every component on the chip is functioning in every cycle. For each workload, we analyze the energy efficiency by summing the consumed energy in each pipeline stage.

\section{Efficient Pipelining}
To better illustrate how intra-layer pipelining, inter-layer pipelining, and batch pipelining work, we define the IFM of the current CNN layer to be $I$, the kernel of the current CNN layer to be $K$, and the OFM of the current CNN layer (also IFM of the next layer) to be $O$. $I$ is a 3D matrix with dimensions $c (channel) \times h (height) \times w (width)$. $K$ is a 4D matrix with dimensions $n (kernel) \times c (channel) \times l (length) \times l (length)$. $O$ is a 3D matrix with dimensions $c (kernel) \times h (height) \times w (width)$.

\subsection{Intra-layer Pipelining}
For the intra-layer pipeline, it takes $h \times w$ logical cycles to pass the entire IFM into this pipeline. Note that that one intra-layer pipeline processes one pixel from all channels. There are four different intra-layer pipelines for one CNN layer depending on whether the layer is mapped to a single tile or multiple tiles and whether the layer has pooling operations at the end. Specifically, single-mapped tile without pooling requires 24 cycles; single-mapped tile with pooling requires 29 cycles; multi-mapped tile without pooling requires 26 cycles; multi-mapped tile with pooling requires 31 cycles.

\subsection{Inter-layer Pipelining}
For inter-layer pipeline, we observe that we don't need to wait for the current layer to produce the entire OFM in order to start the next layer. We only need to wait for enough information from the current layer that is able to start the first convolution of the next layer. The number of values in $O$ that the next layer needs to wait is shown as
\begin{equation}
valuesWait = (w \times (l - 1) + l) \times n
\end{equation} and the number of cycles the next layer needs to wait is shown as
\begin{equation}
cyclesWait = w \times (l - 1) + l
\end{equation} where the kernel strides in the row-majored fashion.

\begin{figure*}[t!]
\centering
\includegraphics[width=0.75\textwidth, height=6.6cm]{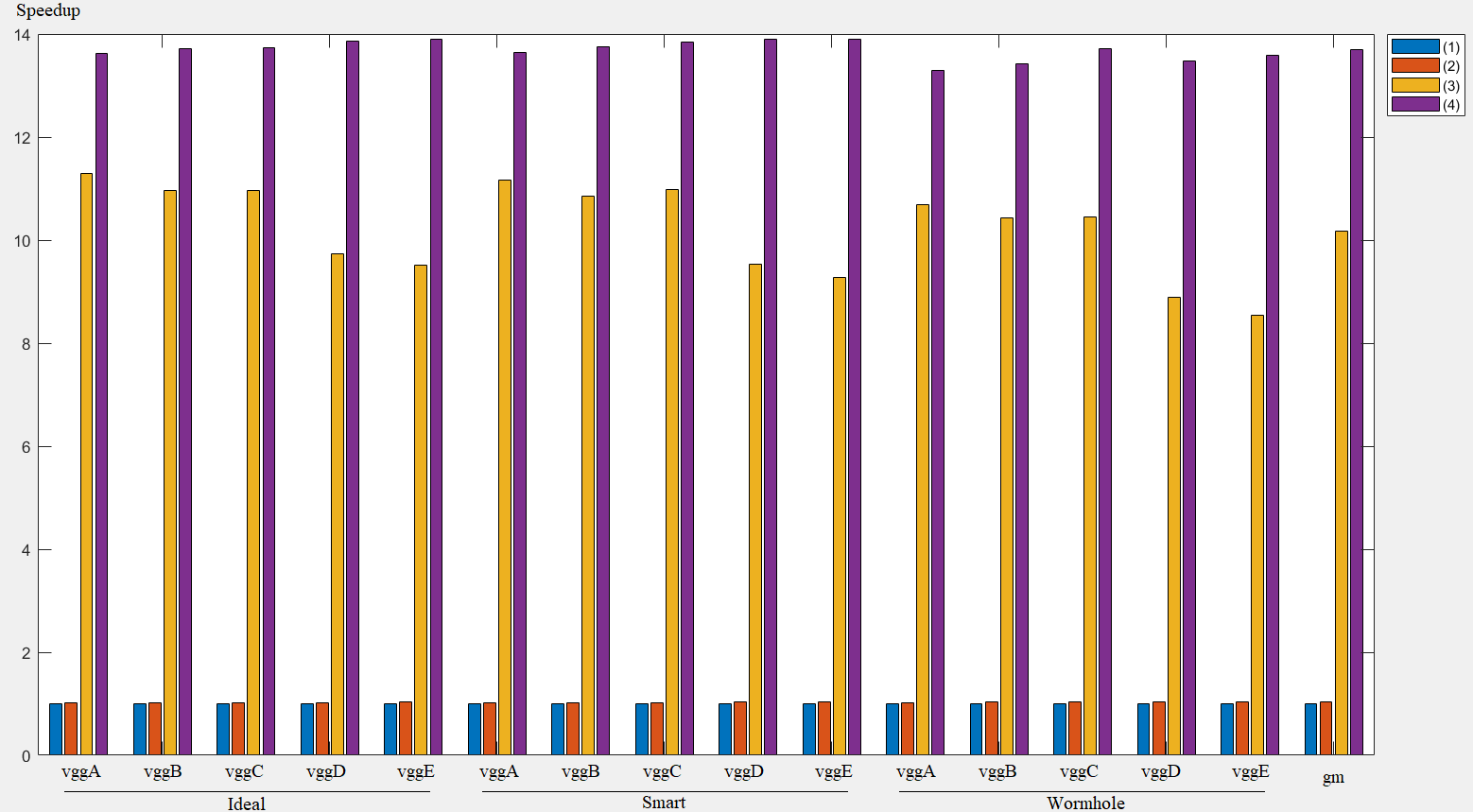}
\caption{Speedup of Each VGG due to Different Pipelinings.}
\label{pipe}
\end{figure*}

\begin{figure*}[t!]
\centering
\includegraphics[width=0.75\textwidth, height=6.6cm]{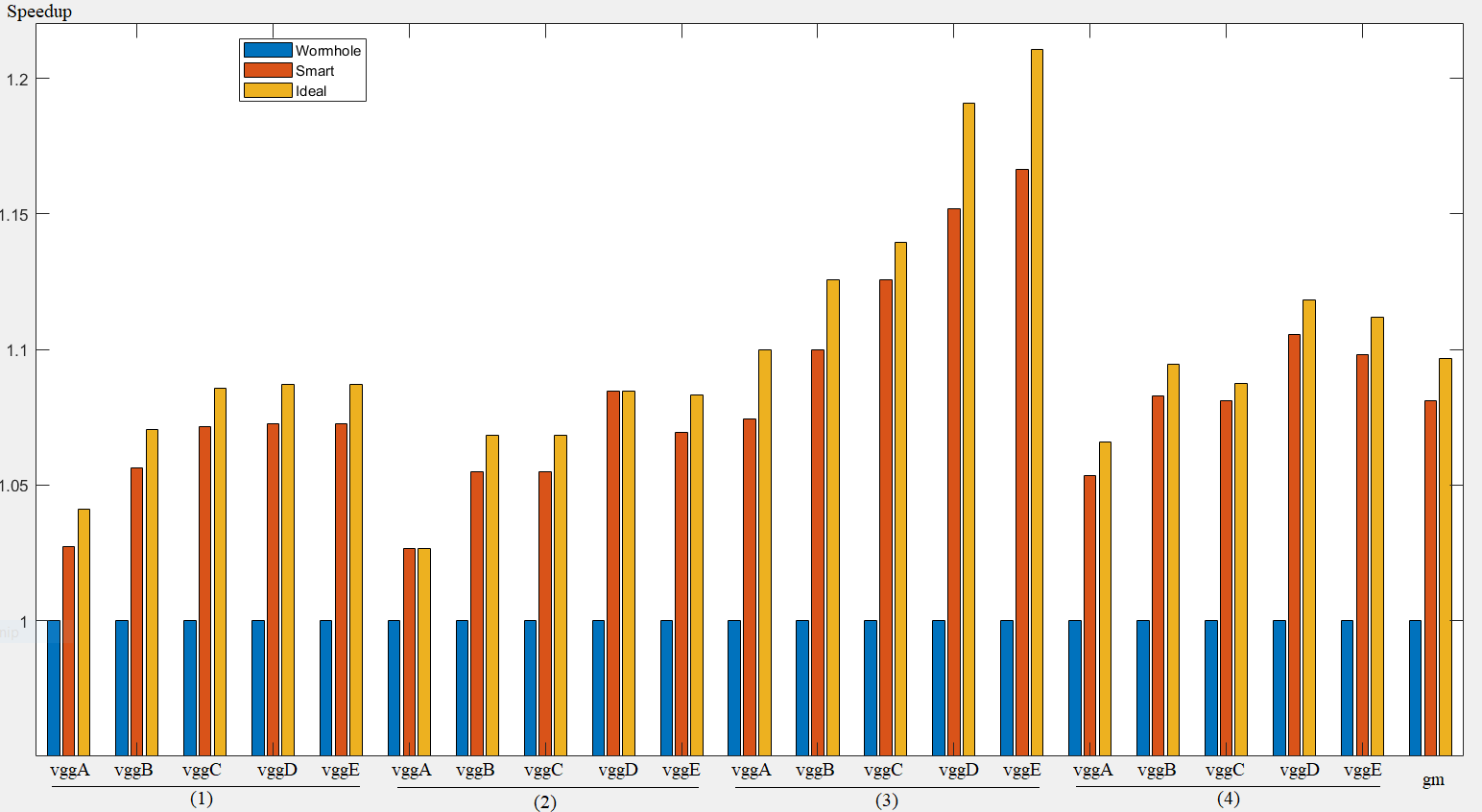}
\caption{Speedup of Each VGG due to Different NoCs.}
\vspace{-10pt}
\label{noc}
\end{figure*}
 
\subsection{Batch Pipelining}
For batch pipelining, we observe that we can design a pipeline that overlaps the latency between input images which come consecutively at a certain rate. We follow two design principles to design the batch pipeline. First, there should be no structural hazard, which means in the same cycle, the pipeline cannot process a specific layer (say layer 1) from two or more different images. In other words, a specific layer in a specific cycle can only process one single image. Second, dependencies between consecutive layer (say layer 1 and layer 2) should be strictly followed for all images. In other ways, if layer 2 has to wait for 2 cycles after layer 1 starts, all layer 2 from all images have to wait for 2 cycles after the corresponding layer 1 starts.

\section{SMART Flow Control}
The topology of a NoC describes the connection between routers via links/channels. NoC topologies include bus, ring, mesh, torus, flattened butterfly, fully connected and so on. The most common topology is a 2D mesh because it can be laid out easily. In our design, the NoC is a $16 \times 20$ 2D mesh topology.

The routing of a NoC describes the links that a flit takes from the source router to the destination router. For example, XY routing means when choosing the routing path from the source to the destination, the flit always goes horizontal (X direction) and then vertical direction (Y direction). In addition, a turn model such as north-last model or east-first model can be used. It disallows some turns to get rid of deadlocks in the NoC. In our design, we use XY routing. In addition, we set the link width to be 128 bits, which is the flit size.

The flow control of a NoC describes when a flit can traverse to the downstream router or it has to stay at the upstream router if there is traffic in the NoC. The most common algorithm is the wormhole flow control. In wormhole, the link is allocated at the packet level and the buffer is allocated at the flit level. It significantly improves the performance of virtual cut-through flow control because its buffer can have flits from different packets. However, wormhole still suffers from poor link utilization and results in HoL (head-of-line) blocking. HoL blocking means if the first flit in the buffer cannot move, all of the rest flits in the buffer cannot move either. In our design, we use wormhole as one baseline.

In order to present the total latency in cycles to send a packet from the source to the destination, we define the wire delay for one link to be $t_{w}$, the hop count to be $H$, the contention delay to be $t_{c}$, and the serialization delay to be $t_{s}$. A typical formula for latency in cycles is defined as
\begin{equation}
T = t_{w} \times H + t_{c} + t_{s}
\end{equation}
where the bottleneck is the term $H$. The ideal solution to reduce $H$ down to 1 is to use the fully connected topology. However, since it's nearly impossible to lay out a topology like this, we resort to smart flow control algorithm which makes the NoC behave closely to an ideal fully connected topology.

Our NoC model uses SMART flow control from \cite{SMART} to reduce NoC latency and increase the overall throughput. Place-and-route repeated wires can go up to 16 mm in 1 ns in 45 nm technology node. It can go further in the projected 32 nm technology node because wire delay remains constant or decreases slightly due to technology scaling. Therefore, on-chip wires can go fast enough to transfer across the chip within 1 or 2 clock cycles. The high level idea to achieve smart flow control is to use multiplexers to bypass the routers on the path from the source to the destination. However, the bottleneck of SMART happens when two different packets are sent in the NoC at the same cycle and the two packets share some common links. Specifically, we need to set two different priorities for the two paths when setting up the SSRs (setup requests) to ensure the correct functionality.

\begin{figure}[t!]
\centering
\includegraphics[width=85mm]{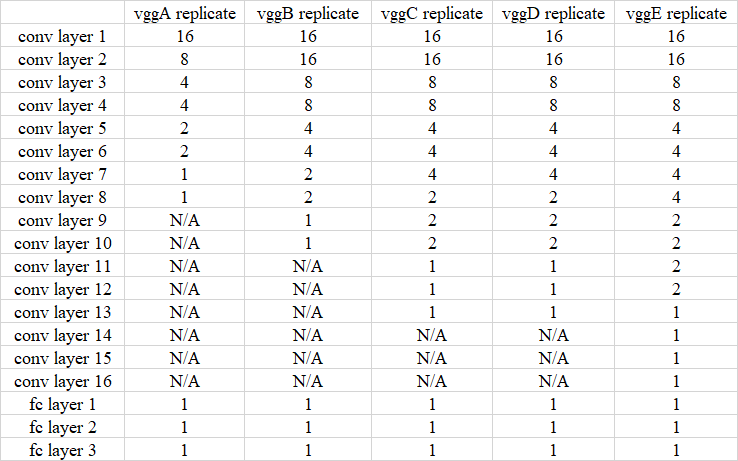}
\caption{Weight Replications of Each VGG.}
\vspace{-15pt}
\label{replication}
\end{figure}

\section{Evaluation}
\subsection{Simulators}
In the experiment, we run the cycle-accurate simulation for the processing side by building a C++ simulator from scratch. We use the cycle-accurate garnet2.0 simulator for the interconnect side. 

\subsection{Workloads and Benchmarks}
We use VGG (A-E) \cite{VGG} for the large-scale data set ImageNet \cite{ImageNet} as our workloads. VGG makes a thorough evaluation of networks of increasing depth using an architecture with very small ($3 \times 3$) convolution filters. Compared to previous CNNs, VGG improves the accuracy of computer vision and pattern recognition tasks by a wide margin, which is achieved by pushing the CNN depth from a few layers to tens of layers. There are a total of $3 \times 4 \times 5 = 60$ benchmarks. There are five different CNNs: VGG (A-E), three different NoCs: ideal, SMART, wormhole, and four different pipelining scenarios: without weight replication and without batch pipelining (1), without weight replication and with batch pipelining (2), with weight replication and without batch pipelining (3), with weight replication and with batch pipelining (4).

\subsection{Weight Replications}
Pooling layers degrade the performance of inter-layer pipelining because the next layer has to wait for the pooled results which come from different columns of the current feature map. This introduces extra pipeline bubbles, increases latency, and decreases throughput. In order to have a more balanced pipeline design, we replicate more weights for the first few layers while replicate less weights for the deep layers. Specifically, all five VGGs are down-sampled five times: $224 \times 224$, $112 \times 112$, $56 \times 56$, $28 \times 28$, $14 \times 14$, $7 \times 7$. Each time a grid of $2 \times 2$ is applied to the whole OFM. In order to satisfy this trend, we also replicate the weights 16 times, 8 times, 4 times, 2 times, and 1 time. Fig. \ref{replication} shows the number of times the weights are replicated in each layer for each VGG. All schemes meet the constraint that there are a maximum of 320 tiles available.

\begin{figure}[t!]
\centering
\includegraphics[width=85mm]{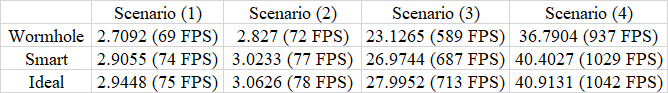}
\caption{VGG E Throughput.}
\vspace{-10pt}
\label{vggE}
\end{figure}

\begin{figure}[t!]
\centering
\includegraphics[width=30mm]{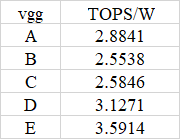}
\caption{Energy Efficiency of Each VGG.}
\vspace{-15pt}
\label{energy}
\end{figure}

\subsection{Results and Analysis}
To explore the effect of different pipelining schemes on the performance, we use scenario (1) as the baseline and calculate the speedup of each scenario by normalizing the throughput to scenario (1). Fig. \ref{pipe} shows the speedup in all four scenarios for each VGG in all three different NoCs. The geometric mean of (2) compared to (1) is $1.0309\times$, (3) compared to (1) is $10.1788\times$, and (4) compared to (1) is $13.6903\times$. Note that for the best pipelining setup in scenario (4), it achieves a speedup close to $16\times$. We don't need to replicate the weights in all layers by 16 times to achieve this speedup. Instead, we replicate weights decreasingly as the layers become deeper and the size of OFM decreases to make a balanced pipeline design. Note that the results in Fig. \ref{pipe} are projected results, which are not directly from garnet2.0.

\begin{figure*}[t!]
\centering
\includegraphics[width=\textwidth, height=10cm]{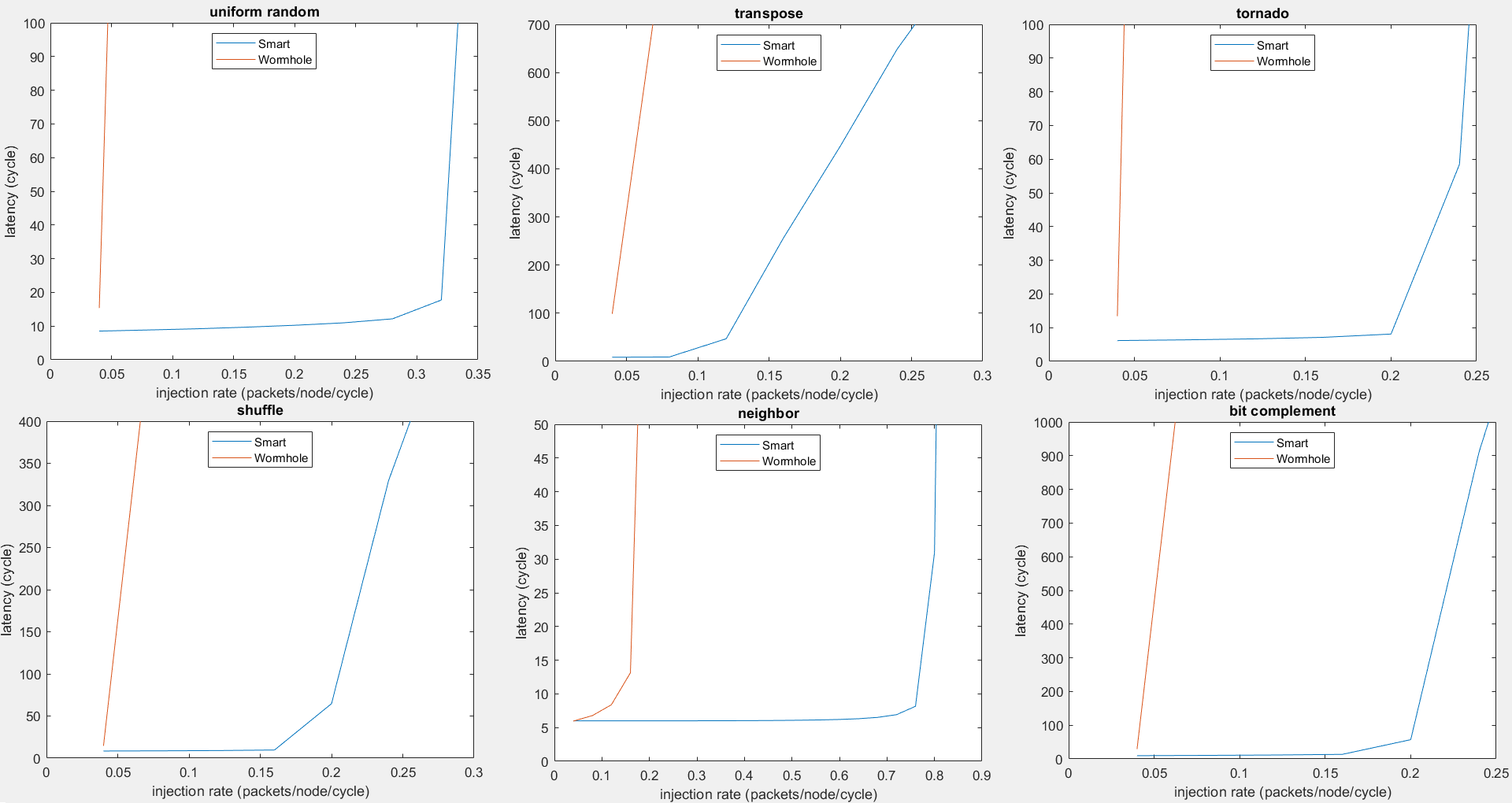}
\caption{Latency Comparison.}
\label{111}
\end{figure*}

\begin{figure*}[t!]
\centering
\includegraphics[width=\textwidth, height=10cm]{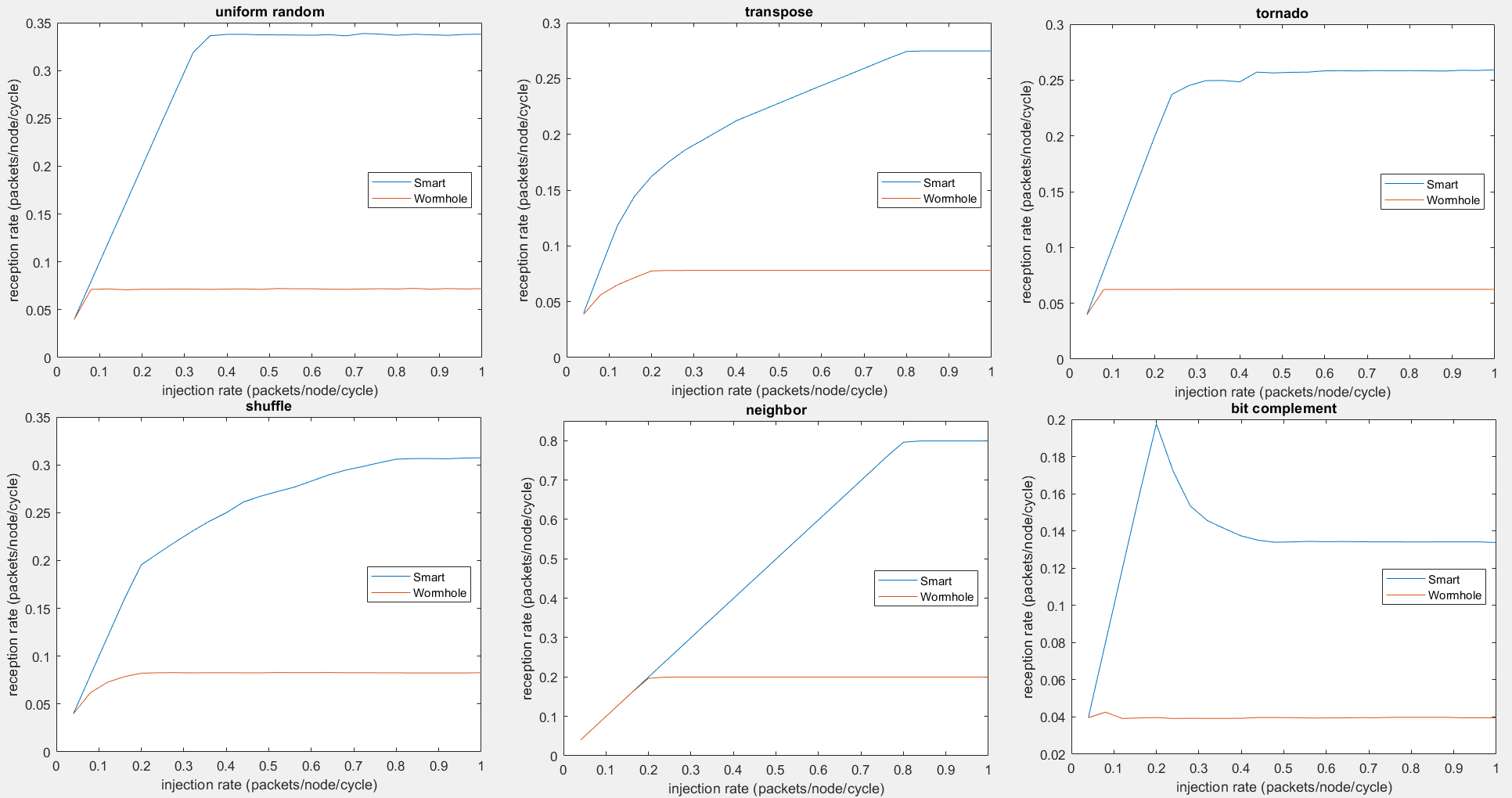}
\caption{Reception Rate Comparison.}
\label{222}
\end{figure*}

To explore the effect of different NoCs on the performance, we use wormhole as the baseline and calculate the speedup of all three NoCs (ideal, SMART, wormhole) by normalizing the throughput to wormhole. Fig. \ref{noc} shows the speedup in all three NoCs for each VGG in all four pipelining scenarios. The geometric mean of ideal compared to wormhole is $1.0809\times$ and SMART compared to wormhole is $1.0965\times$. Note that SMART NoC achieves better speedup for more aggressive pipelining because more aggressive pipelining has heavier traffic in the NoC, so the performance of SMART NoC improves effectively while the performance of wormhole NoC degrades performance even more. Note that the results in Fig. \ref{noc} are projected results, which are not directly from garnet2.0.

The best throughput is achieved when running VGG E. Fig. \ref{vggE} shows the throughput and the corresponding frame rate of the architecture when running VGG E in all combinations of flow controls and pipelining scenarios. We also report the energy efficiency of the architecture when processing each VGG, as shown in Fig. \ref{energy}. Note that weight replications, batch pipelining, and different flow control algorithms don't affect energy efficiency much because with the total amount of energy consumed depends mostly on the amount of operations in the workload.

\section{Evaluation of SMART Flow Control using Synthetic Traffic}
\subsection{Simulation Setup}
We use garnet2.0 to evaluate the performance of SMART flow control compared to wormhole flow control using six synthetic traffics including uniform random, transpose, tornado, shuffle, neighbor, and bit complement. In addition, we set up the two flow controls using an $8 \times 8$ mesh and XY routing algorithm. For SMART flow control, we assume $HPCmax \geq 14$, since the wire delay for a 1 $mm^2$ chip can be taken care of within 1 clock cycle \cite{SMART}.

\subsection{Results and Analysis}
Fig. \ref{111} shows the injection rate vs latency plot for all six synthetic traffics. For uniform random, transpose, tornado, shuffle, and bit complement, wormhole saturates when the injection rate is around 0.05 while SMART saturates when the injection rate is around 0.25. For neighbor, wormhole saturates when the injection rate is around 0.2 while SMART saturates when the injection rate is around 0.8. It's obvious that SMART has higher throughput than wormhole.

Fig. \ref{222} shows the injection rate vs reception rate plot for all six synthetic traffics. For uniform random, transpose, tornado, and shuffle, wormhole saturates when the reception rate is around 0.07 while SMART saturates when the reception rate is around 0.3. For neighbor, wormhole saturates when the reception rate is around 0.2 while SMART saturates when the reception rate is around 0.8. For bit complement, wormhole saturates when the reception rate is around 0.04 while SMART saturates when the reception rate is around 0.14. It's obvious that SMART has higher reception rate than wormhole.

\section{Conclusion}
In this paper, we propose a ReRAM-based PIM architecture for fast and efficient CNN inference. Our optimizations come from three aspects. First, we design intra-layer pipelining, inter-layer pipelining, and batch pipelining to exploit parallelism in the architecture and increase overall throughput. Second, we optimize the performance of the NoC using SMART flow control. Third, we leverage weight replications to maximize parallelism and further accelerate the architecture. Our simulation shows the different pipelining and weight replications achieves a speedup of $14\times$ compared to the baselines from the processing side. SMART flow control achieves a speedup of $1.08\times$ compared to the baselines from the interconnect side. Our evaluation of SMART flow control using synthetic traffics show that SMART outperforms wormhole by a wide margin in the communication-heavy synthetic traffics. Since NoC only represents a small portion of performance and power within the proposed PIM architecture, SMART flow control enhances the overall performance by a small margin while most speedup is achieved from the processing side by designing efficient pipelining and leveraging weight replications.

\end{document}